\documentstyle[12pt,epsfig]{ws-art}
\newcommand{\beq}{\begin{equation}}
\newcommand{\eeq}{\end{equation}}
\newcommand{\beqa}{\begin{eqnarray}}
\newcommand{\eeqa}{\end{eqnarray}}
\newcommand{\nn}{\nonumber}

\newcommand{\Dis}[1]{$\displaystyle #1$}

\newcommand{\hc}{\mbox{{\rm h.c.}}}
\newcommand{\MNS}{M_2}
\newcommand{\mlN}{M_1}
\newcommand{\Mu}{m_{{ u}}}
\newcommand{\Au}{A_{{ u}}}
\newcommand{\Md}{m_{{ d}}}
\newcommand{\Ad}{A_{{ d}}}

\newcommand{\suq}{\tilde{{ u}}}
\newcommand{\Tsp}{\mbox{\scriptsize T}}
\newcommand{\eps}{\epsilon}
\newcommand{\sdq}{\tilde{{ d}}}

\newcommand{\thW}{\theta_{\rm W}}
\newcommand{\mW}{m_{{ W}}}
\newcommand{\MsQU}{\wtilde{3}{0.8}{M}_{\hspace*{-1mm}U}}

\newcommand{\Msu}{\wtilde{3}{0.2}{m}_{U}}
\newcommand{\Msd}{\wtilde{3}{0.2}{m}_{\!D}}

\newcommand{\Ms}{\wtilde{3}{0.8}{m}}
\newcommand{\wtilde}[3]{\settowidth{\ltT}{\Dis{#3}}
\makebox[\ltT]{$\rule{#2\mmh}{0mm}
\widetilde{\makebox[#1\mm]{\Dis{#3\rule{#2\mm}{0mm}}}}$}}

\newlength{\ltT}
\newlength{\mmh}
\newlength{\mm}
\def\gsim{\mathrel{\rlap{\raise 1.5pt \hbox{$>$}}\lower 3.5pt
\hbox{$\sim$}}}
\def\lsim{\mathrel{\rlap{\raise 1.5pt \hbox{$<$}}\lower 3.5pt
\hbox{$\sim$}}}
\def\GeV{{\rm GeV}}
\def\TeV{{\rm TeV}}
\def\Order{{\cal O}}
\def\slash#1{#1 \hskip -0.5em /}

\newcounter{figureno}
\newenvironment{capt}{
\phantom{mmmm}
\vspace*{10mm}
\parindent=0pt
\addtocounter{figureno}{1}

\begin{minipage}[t]{150mm}
\small\sl Figure~\thefigureno.\ }{\end{minipage}
\vspace*{-5mm}}

\begin{document}

\def\Month{\ifcase\month\or
January\or February\or March\or April\or May\or June\or 
July\or August\or September\or October\or November\or December\fi}
\def\slash#1{#1 \hskip -0.5em /}
%

\title{
LIGHTEST MSSM HIGGS BOSON PRODUCTION AND ITS TWO-PHOTON DECAY AT 
THE LHC\thanks{To appear in:
{\it Proceedings of the Workshop: Quantum Systems: 
New Trends and Methods}, Minsk, Belarus,
June 3--7, 1996; {hep-ph/9608315} }}
\author{
      B. KILENG \\
{\it NORDITA, Blegdamsvej 17, DK-2100 Copenhagen \O, Denmark} \\
\\
      P. OSLAND \\
{\it University of Bergen, All\'egt.~55, N-5007 Bergen, Norway}
\and
      P.N. PANDITA \\
{\it North Eastern Hill University, Permanent Campus, 
Shillong 793022, India}}
\maketitle

\begin{abstract}
We present an analysis of the production and two-photon decay 
of the lightest CP-even Higgs boson of the Minimal Supersymmetric 
Standard Model (MSSM) at the Large Hadron Collider (LHC).
A rather general model is considered, without supergravity
constraints.
All parameters of the model are taken into account, we especially
study the dependence of the cross section on the squark masses,
on the bilinear parameter $\mu$ and the trilinear
supersymmetry breaking parameter $A$.
Non-zero values of these parameters lead to significant mixing
in the squark sector, and, thus, affect the masses of Higgs bosons
through radiative corrections,
as well as their couplings to squarks.
The cross section times the two-photon branching ratio 
of $h^0$ is of the order of 15--25~fb in much
of the parameter space that remains after imposing the
present  experimental constraints on the parameters.
\end{abstract}

\section{Introduction}
The most important production mechanism for the neutral SUSY Higgs 
bosons\cite{MSSM} at the Large Hadron Collider (LHC) is the gluon
fusion mechanism, $pp\to gg\to h^0$, $H^0$, $A^0$,\cite{GGMN}
and the Higgs radiation off top and bottom quarks.\cite{Kunsztetal}
Except for the small range in the parameter space where the
heavy neutral Higgs $H^0$ decays into a pair of $Z$ bosons,
the rare $\gamma\gamma$ decay mode, apart from $\tau\tau$ decays,
is a promising mode to detect the neutral Higgs particles,
since $b$ quarks are hard to separate from the QCD background. 
It has been pointed out that the lightest Higgs 
could be detected in this mode for sufficiently large values 
of the mass of the pseudoscalar Higgs boson 
$m_A\gg m_Z$.\cite{KunsztZ,Baer}

Here we present results of a recent study\cite{KOP} 
of the hadronic production
and subsequent two-photon decay of the lightest $CP$-even Higgs boson
($h^0$) of the MSSM,
which is valid for the LHC energy of $\sqrt{s}=14~\TeV$, 
focussing on the case of intermediate-mass squarks.
The case of heavier squarks has been discussed
else\-where.\cite{KOP,KOPmos}
A related study has been presented by Kane et al.\cite{Kane}
They consider a model where parameters are related by supergravity,
but otherwise chosen randomly within their allowed ranges.
As mentioned, the gluon fusion mechanism is the dominant
production mechanism of SUSY Higgs bosons in high-energy
$pp$ collisions throughout the entire Higgs mass range.
We study the cross section for the production of the $h^0$
and its decay, taking into account all 
the parameters of the Supersymmetric Standard Model.
In particular, we discuss the dependence on the squark masses,
and take into account the mixing in the squark sector, the chiral mixing.
This also affects the Higgs boson masses 
through appreciable radiative corrections,
and was previously shown to lead to large corrections to 
the rates.\cite{Kileng}

In the calculation of the production of the Higgs through
gluon-gluon fusion, we include in the triangle graph
all the squarks, as well as $b$ and $t$ quarks, the lightest quarks
having a negligible coupling to the $h^0$.
On the other hand, in the calculation of decay of the Higgs to two
photons, we include in addition to the above, all the sleptons,
$W^\pm$, charginos and the charged Higgs boson.

An important role is played in our analysis by the bilinear
Higgs coupling $\mu$, which occurs in the Lagrangian through the term
\begin{equation}
{\cal L}
= \biggl[
    - \mu \hat{H}^{\Tsp}_{1} \eps \hat{H}_{2}\biggr]_{\theta\,\theta}
    + \hc,
\end{equation}
where $\hat{H}_{1}$ and $\hat{H}_{2}$ are the Higgs superfields with
hypercharge $-1$ and $+1$, respectively.
Furthermore, the Minimal Supersymmetric Model contains several 
soft super\-sym\-metry-breaking terms. 
We write the relevant soft terms in the Lagrangian 
as follows\cite{GirGri}
\begin{eqnarray}
\label{EQU:Lagrangefour}
{\cal L}_{\mbox{{\scriptsize Soft}}}
& = & \Biggl\{ \frac{g\Md\Ad}{\sqrt{2}\;\mW\cos\beta}
      Q^{\Tsp}\eps H_{1}\sdq^{R}
    - \frac{g\Mu\Au}{\sqrt{2}\;\mW\sin\beta}
      Q^{\Tsp}\eps H_{2}\suq^{R}
    + \hc \Biggr\} \nonumber \\
& & - \MsQU^{2}Q^{\dagger}Q - \Msu^{2}\suq^{R\dagger}\suq^{R}
    - \Msd^{2}\sdq^{R\dagger}\sdq^{R}
    - M_{\!H_{1}}^{2} H_{1}^{\dagger}H_{1}
    - M_{\!H_{2}}^{2} H_{2}^{\dagger}H_{2} \nn \\*
& & + \frac{\mlN}{2}\left\{\lambda\lambda +\bar{\lambda}\;
      \bar{\lambda}\right\}
    + \frac{\MNS}{2} \sum_{k=1}^{3} \left\{\Lambda^{k}\Lambda^{k}
    + \bar{\Lambda}^{k}\bar{\Lambda}^{k}\right\},
\end{eqnarray}
(see also ref.~[9])
with subscripts $u$ (or $U$) and $d$ (or $D$) referring to up and
down-type quarks.
The Higgs production cross
section and the two-photon decay rate depend significantly 
on several of these parameters.

The Higgs production cross section and the two-photon decay rate 
depend on the gaugino and squark masses, the latter being determined
by, apart from the soft-supersymmetry breaking trilinear coefficients
($A_u$, $A_d$) and the Higgsino mixing parameter $\mu$, the soft
supersymmetry-breaking masses, denoted in 
eq.~(\ref{EQU:Lagrangefour})
by $\MsQU$, $\Msu$ and $\Msd$, respectively.
For simplicity, we shall consider the situation where
$\MsQU=\Msu=\Msd\equiv\Ms$, with $\Ms$ chosen to be
150~GeV for the first two generations,\footnote{Increasing 
$\Ms$ to 500 GeV for the first two generations leads
to an increase in the cross section times the two-photon
decay rate of about 5\%.}
and varied over the values 
150, 250, 500 and 1000~GeV for the third generation.
As discussed in ref.~[6], it suffices to consider $A$ positive
and vary the sign of $\mu$.

In Sec.~2,  we discuss the implications of the nonzero 
values of $A$ and $\mu$ on the Higgs masses, together with the
constraints related to the other relevant masses. We then go on 
to study the cross sections and decay rates for the lighter 
CP-even Higgs boson in Sec.~3. 

\section{Constraints on the Parameter Space}
\label{sec:limit}
In this section we describe in detail the parameter space relevant for 
the production and decay of the lightest Higgs boson at LHC, and the 
theoretical and experimental constraints on it before presenting
cross sections and decay rates.

At the tree level, the masses of the $CP$-even neutral Higgs 
bosons are given by ($m_{h^0}\le m_{H^0}$)\cite{HHG}
\begin{equation}
m^2_{H^0,h^0}=\frac{1}{2}\left[m_A^2+m_Z^2
\pm\sqrt{\left(m_A^2+m_Z^2\right)^2-4m_Z^2 m_A^2\cos^22\beta}\right],
\end{equation}
which are controlled by two parameters, $m_A$ (the mass of 
the $CP$-odd Higgs boson) and 
$\tan\beta$ ($=v_2/v_1$, where $v_2$ and $v_1$ are the vacuum
expectation values of the two Higgs doublets).
Indeed, the entire Higgs sector at the tree level can be described 
in terms of these two parameters alone.
At the tree level, the mass of the lightest Higgs boson
($m_{h^0}$) is bounded by $m_Z$.
There are, however, substantial radiative corrections\cite{Okada}
to the $CP$-even neutral Higgs masses. 
The radiative corrections are, in general, positive, and they shift
the mass of the lightest neutral Higgs boson upwards.\cite{Okada}
More recent radiative corrections to the Higgs sector\cite{Carena}
which are valid when the squark masses are
of the same order of magnitude, 
have not been taken into account in our study.\cite{KOP}
As long as the ``loop particles" are far from threshold for
real production, the cross section does not depend very strongly
on the exact value of the Higgs mass.

In order to simplify the calculations,
we shall assume that all the trilinear couplings are equal,
\beq
\Au=\Ad\equiv A.
\eeq
Furthermore, we shall take the top-quark mass to be 175~GeV\cite{CDF} 
in our numerical calculations.
The parameters that enter the neutral $CP$-even Higgs mass
matrix are varied in the following ranges:
\beqa
50~\GeV&\leq& m_A \leq 1000~\GeV, \qquad 1.1\leq\tan\beta\leq 50.0,
\nn \\
\qquad 50~\GeV&\leq&|\mu|\leq1000~\GeV, \qquad
0\leq A\leq 1000~\GeV.
\eeqa
These values cover essentially the entire physically interesting
range of parameters in the MSSM.
However, not all of the above parameter values are allowed
because of the experimental constraints on the squark, chargino
and $h^0$ masses.
For low values of $\Ms$, the lightest squark tends to be
too light (below the most rigorous experimental bound of
$\sim 45-48~\GeV$\cite{squarklimit}), or even unphysical
(mass squared negative).
\begin{figure}[htb]
\begin{center}
\setlength{\unitlength}{1cm}
\begin{picture}(15.0,8.0)
\put(0.,-2.0)
{\mbox{\epsfysize=10.0cm\epsffile{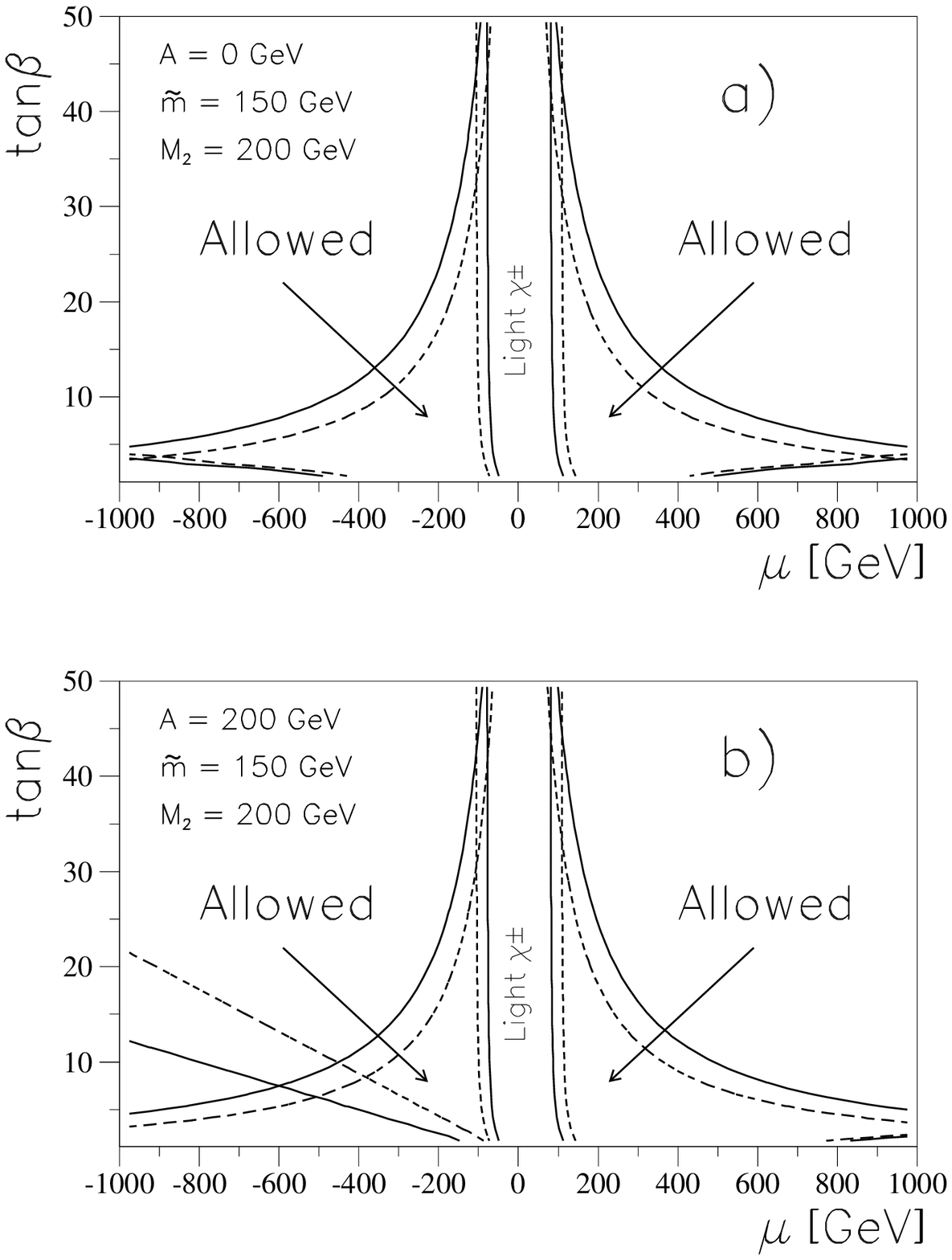}}\hspace*{-8mm}
 \mbox{\epsfysize=10.0cm\epsffile{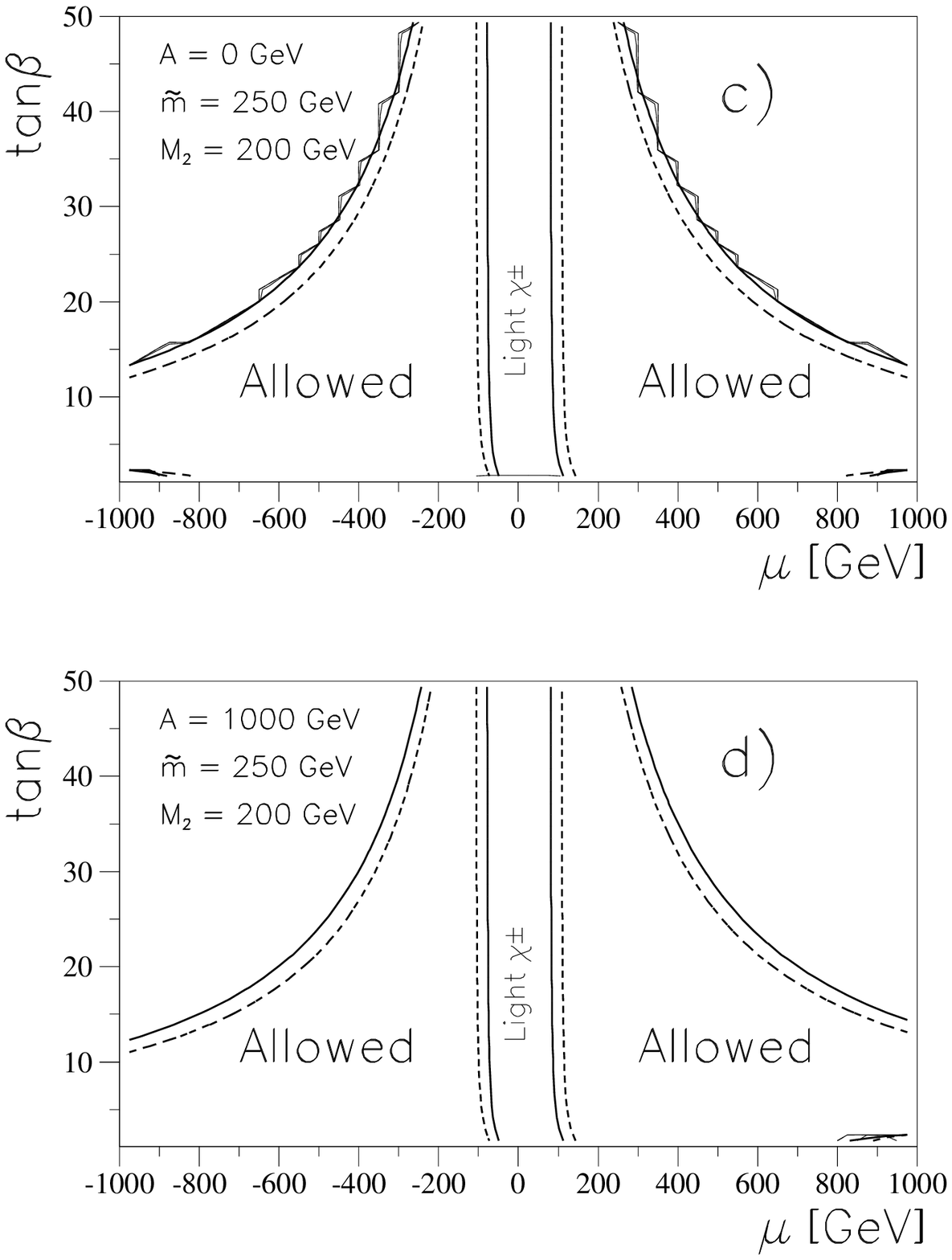}}}
\end{picture}
\vspace*{10mm}
\begin{capt}
Regions in the $\mu$-$\tan\beta$ plane which are ruled out
by too light chargino ($\chi^\pm$) and squark masses.
The gaugino mass scale is $\MNS=200~\GeV$ and $m_A=200~\GeV$.
The solid (dashed) contours for small $|\mu|$ refer to the chargino mass
$m_{\chi^\pm}=68$ $(90)~\GeV$.
Two values of $\Ms$ are considered, {\em left}: $\Ms=150~\GeV$,
{\em right}: $\Ms=250~\GeV$.
For each value of $\Ms$,
two values of the trilinear mixing parameter are considered.
For $\Ms=150~\GeV$, the squark masses are too light or
unphysical in much of the $\mu$-$\tan\beta$ plane.
The hyperbola-like contours give regions that are excluded
by the lightest $b$ squark being below $45~\GeV$ (solid) or
$90~\GeV$ (dashed).
The more straight contours at large $\mu$ and small $\tan\beta$
similarly indicate regions that are excluded by the
lightest $t$ squark.
\end{capt}
\end{center}
\end{figure}
The excluded region of the parameter space is shown
in the left part of fig.~1 for $\Ms=150~\GeV$, $\MNS=200~\GeV$, 
$m_A=200~\GeV$ and for two values of the trilinear coupling $A$.
The allowed region in the $\mu-\tan\beta$ plane decreases 
with increasing $A$, but the dependence on $\MNS$ and $m_A$ 
in this region is rather weak.
In order to have acceptable $b$-squark masses,
$\mu$ and $\tan\beta$ must lie {\it inside} of the
hyperbola-shaped curves. Similarly, in order to have
acceptable $t$-squarks, the corners at large $|\mu|$ and
small $\tan\beta$ must be excluded.


The chargino masses are, at the tree level, given by the expression
\beqa
m^2_{\chi^\pm}
&=& \frac{1}{2}(\MNS^2+\mu^2)+\mW^2 \nn \\
& & \pm\left[\frac{1}{4}(\MNS^2-\mu^2)^2 +\mW^4\cos^2{2\beta}
+\mW^2(\MNS^2+\mu^2+2\mu\MNS\sin{2\beta})\right]^{1/2}.
\eeqa
When $\mu=0$, we see that, for $\tan\beta\gg1$,
the lightest chargino becomes massless.
Actually, small values of $\mu$ are unacceptable
for all values of $\tan\beta$.
The lowest acceptable value for $|\mu|$ will depend
on $\tan\beta$, but that dependence is rather weak.
The excluded region due to the chargino being too light,
increases with decreasing values of $\MNS$.
We note that the radiative corrections to the chargino masses are small
for most of the parameter space.\cite{Lahanas}
In fig.~1 we show the contours in the $\mu$-$\tan\beta$
plane outside of which the chargino has an acceptable mass 
($>68~\GeV$)\cite{chino}.
By the time the LHC starts
operating, one would have searched for charginos with
masses up to 90~GeV at LEP2. Contours relevant for LEP2
are also shown.


For $\Ms=250~\GeV$ (right part of fig.~1),
the region excluded due to low or unphysical squark masses 
is much reduced, and at $\Ms=500~\GeV$ it is practically 
absent.\cite{KOP}
However, for larger values of $\Ms$ the experimental constraints
on the $h^0$ mass\cite{LEPdata} rule out some corners 
of the $\mu-\tan\beta$ plane.\cite{KOP}
The extent of these forbidden regions in the parameter space
grow rapidly as $m_A$ decreases below $\Order(150~\GeV)$.
They also increase with increasing values of $A$.

As discussed above,
the mass of the lighter $CP$-even Higgs boson $h^0$ will depend
significantly on $\mu$, $\tan\beta$, $A$ and $\Ms$, through
the radiative corrections. For $\Ms=250~\GeV$, and two values
each of $m_A$ (100 and 200~GeV) and $A$ (0 and 1000~GeV),
the dependence on $\mu$ and $\tan\beta$ is displayed in fig.~2.
At large $|\mu|$ and large $\tan\beta$,
the radiative corrections are large and negative, driving the value of
$m_{h^0}$ well {\it below} the tree-level value.
(Those regions practically coincide with those where the $b$ squarks
are very light or unphysical.)
\begin{figure}[thb]
\begin{center}
\setlength{\unitlength}{1cm}
\begin{picture}(15.0,10.0)
\put(3.3,-2)
{\mbox{\epsfysize=12cm\epsffile{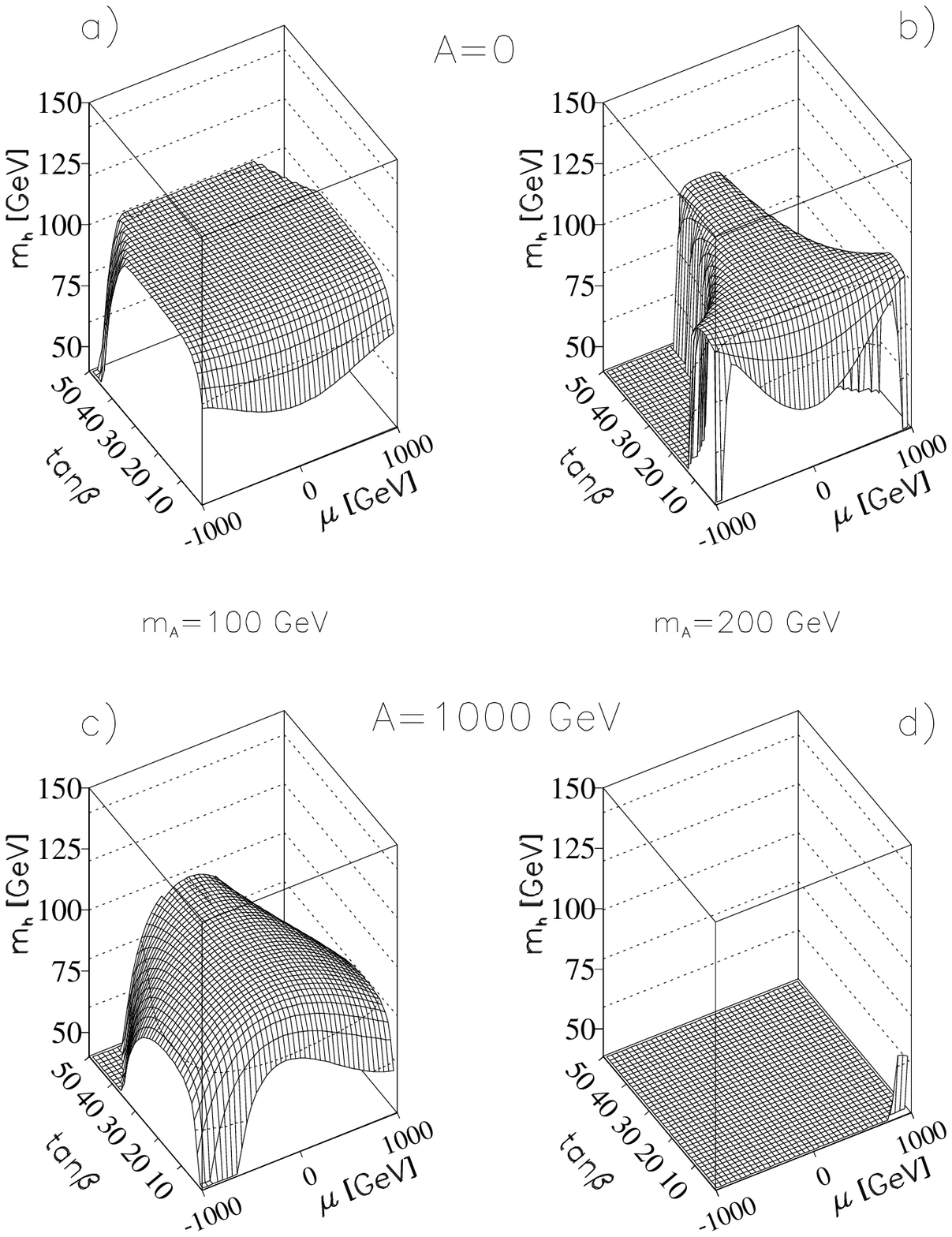}}}
\end{picture}
\begin{capt}
Mass of the lightest $CP$-even Higgs boson vs.\ $\mu$
and $\tan\beta$, 
for $\MNS=200~\GeV$ and $\Ms=250~\GeV$.
Two values of $m_A$ and two values of $A$ are considered: 
a)~$m_A=100~\GeV$, $A=0$,
b)~$m_A=200~\GeV$, $A=0~\GeV$, 
c)~$m_A=100~\GeV$, $A=1000~\GeV$,
d)~$m_A=200~\GeV$, $A=1000~\GeV$.
\end{capt}
\end{center}
\end{figure}


The charged Higgs boson mass is given by
\beq
m^2_{H^\pm}=\mW^2+m_A^2+\Delta,
\eeq
where $\Delta$ arises due to radiative corrections
and is a complicated function of the parameters of 
the model.\cite{Brignole}

The radiative corrections to the charged Higgs mass are, in general,
not as large as in the case of neutral Higgs bosons.
In certain regions of parameter space the radiative corrections can,
however, be large.
This is the case when the trilinear mixing parameter $A$ is
large, $m_A$ is small, and when furthermore $\tan\beta$ is large.
We shall include the effects of non-zero $A$ and $\mu$
in the calculation of the charged Higgs mass.
The present experimental lower bound of 40--45~GeV\cite{chargedH} 
on the charged Higgs is not restrictive,
but presumably by the time the LHC starts operating, one will 
have searched for charged Higgs bosons at LEP2 with mass up to 
around 90~GeV.
Even this bound does not appreciably restrict the parameter space.
                    

The neutralino mass matrix depends on the four parameters
$\MNS$, $\mlN$, $\mu$ and $\tan\beta$.
However, one may reduce the number of parameters by assuming that 
the MSSM is embedded in a grand unified theory so that the
SUSY-breaking gaugino masses are equal to
a common mass at the grand unified scale.
At the electroweak scale, we then have\cite{Inoue}
$\mlN=(5/3)\tan^2\thW\, \MNS$.
We shall assume this relation throughout in what follows.
The neutralino masses enter the calculation through the total
width of the Higgs boson.
We here present numerical results for the gaugino mass parameter
being $\MNS=200~\GeV$.
A wider range of values is considered in ref.~[6].
The experimental constraints on the lightest
neutralino mass rule out regions of the parameter space
similar to those ruled out by the charginos.\cite{neutr}

\section{Cross Section and Two-Photon Decay}
\label{sec:Xsects-h}

The cross section for
$pp\to h^0 X$,
is calculated from the triangle diagram convoluted with
the gluon distribution functions,
\begin{equation}
\sigma
=\sqrt2\,\pi\,G_{\rm F}\,
\biggl(\frac{\alpha_s}{8\pi}\biggr)^2 \, \frac{m_{h^0}^2}{s}
\Big|\sum_{k} I_k(\tau)\Big|^2 
\int_{-\log(\sqrt s/m_{h^0})}^{\log(\sqrt s/m_{h^0})} {\rm d} y \,
G\Bigl(\frac{m_{h^0}}{\sqrt s}\,e^y\Bigr)\,
G\Bigl(\frac{m_{h^0}}{\sqrt s}\,e^{-y}\Bigr),
\end{equation}
with contributions from various diagrams ($k$). For the standard 
case of a top-quark loop,
\begin{equation}
I(\tau)=\frac{\tau}{2}\biggl\{1-(\tau-1)
\biggl[{\rm arcsin}\biggl(\frac{1}{\sqrt \tau}\biggr)\biggr]^2\biggr\},
\nonumber
\end{equation}
and $\tau=(2m_t/m_{h^0})^2>1$.

For $\MNS=200~\GeV$, $\Ms=500~\GeV$, and $\mu=500~\GeV$, 
plots are given in refs.~[6,7].
The following features are noteworthy:
\begin{itemize}
\item The cross section decreases appreciably for large values of $A$.
This is mainly due to an increase in the $h^0$ mass.
\item The cross section increases sharply for small values of
$\tan\beta$, and also at small $m_A$.
The increase at small $\tan\beta$ is caused by the $h^0$ becoming light.
At small values of $m_A$ and large $A$, the couplings of $h^0$
to $b$ quarks and $\tau$ leptons become large, making
the cross section very large in this region.
\end{itemize}

\begin{figure}[thb]
\begin{center}
\setlength{\unitlength}{1cm}
\begin{picture}(15.0,5.5)
\put(-0.2,-1)
{\mbox{\epsfysize=5.5cm\epsffile{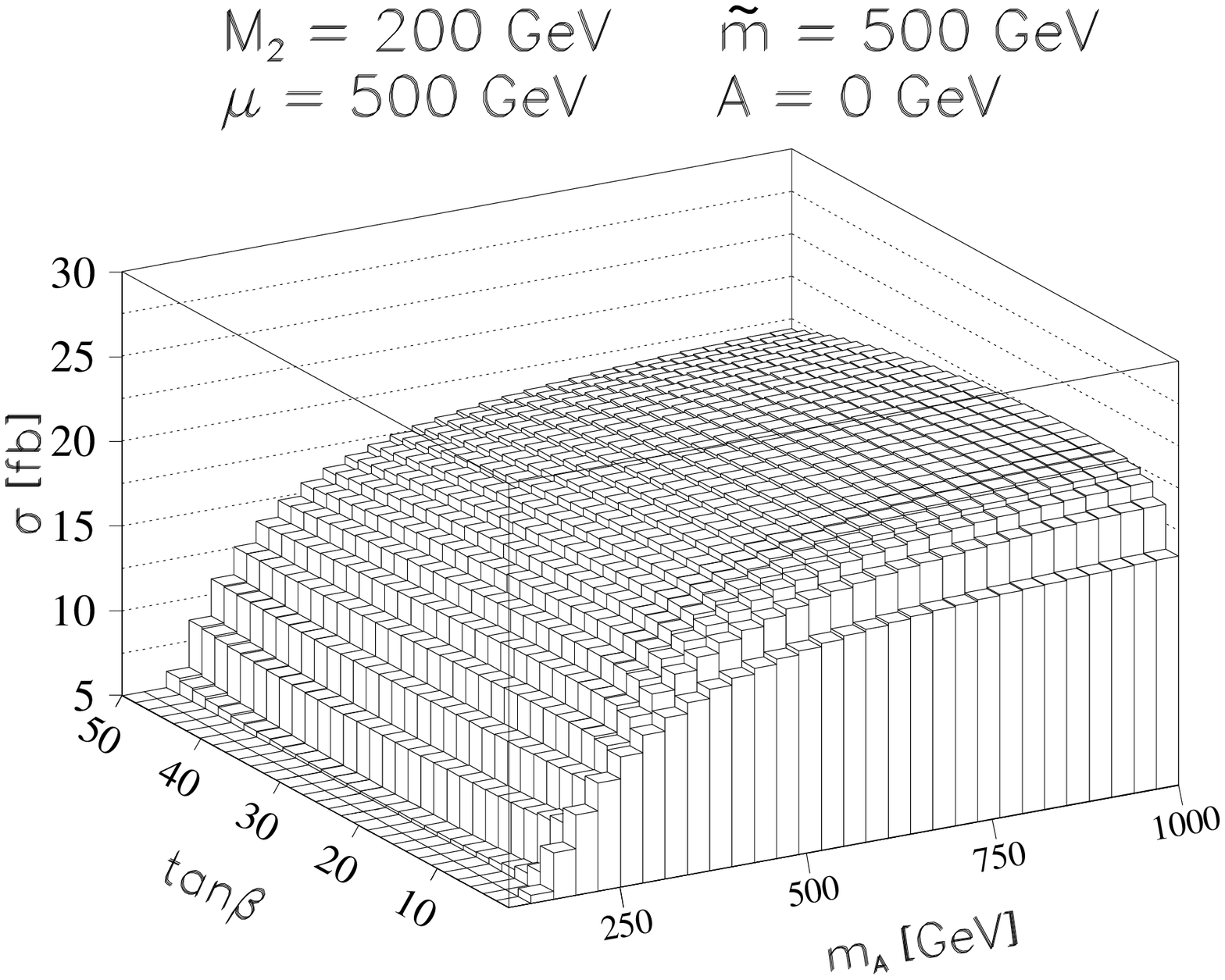}} 
 \mbox{\epsfysize=5.5cm\epsffile{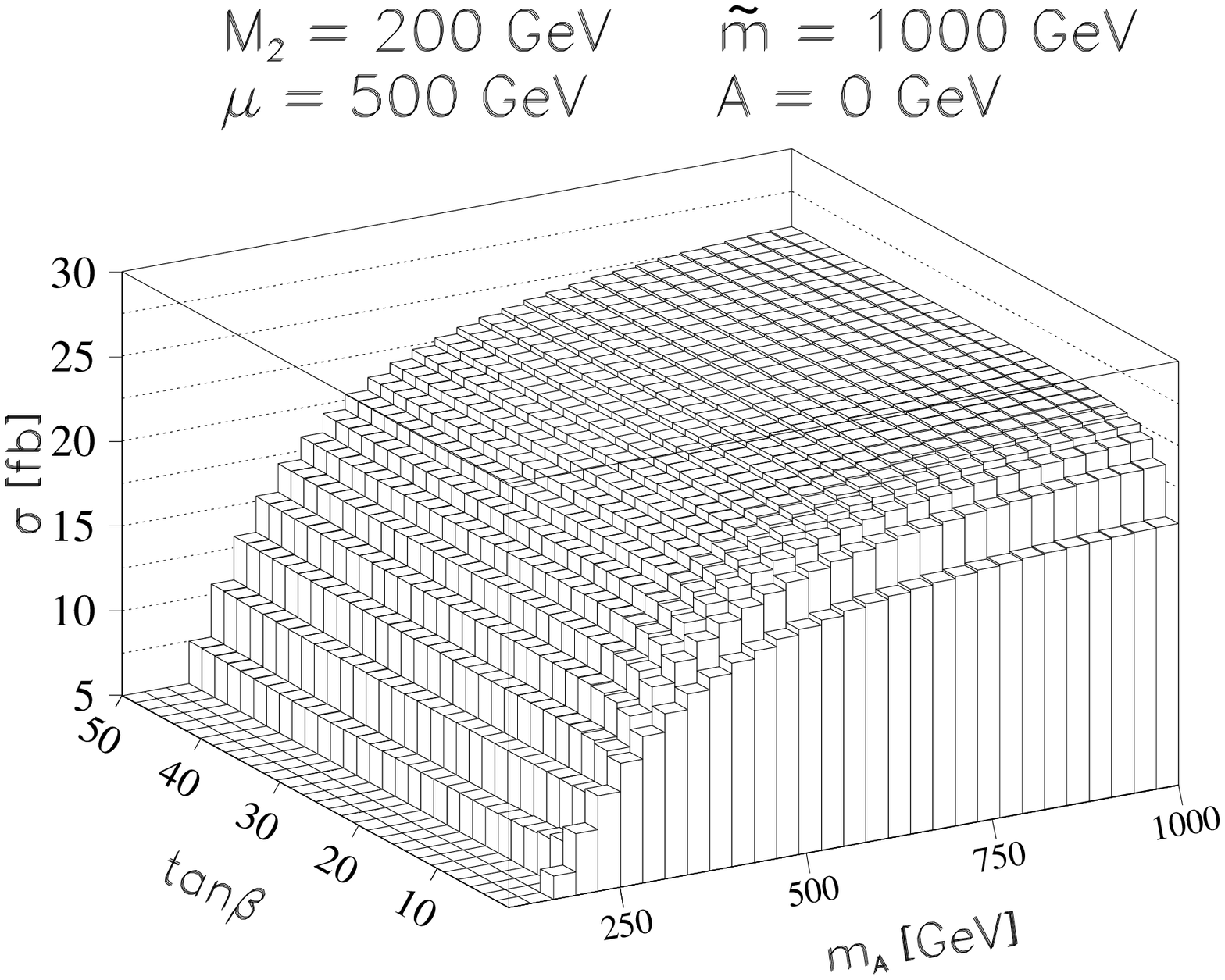}}}
\end{picture}
\begin{capt}
Cross section for $pp\to h^0X\to\gamma\gamma X$
as a function of $m_A$ and $\tan\beta$
for $\MNS=200~\GeV$,
two values of $\Ms$:
{\em left}: $\Ms=500~\GeV$, {\em right}: $\Ms=1000~\GeV$.
For both cases, $\mu=500~\GeV$, and $A=0$.
\end{capt}
\end{center}
\end{figure}
The two-photon decay rate is found\cite{KOP,KOPmos} to increase 
sharply at large values of $A$, but this does not result in 
a correspondingly larger rate for the process
\beq
\label{eq:pp-h0-gammagamma}
pp\to h^0 X \to \gamma\gamma X,
\eeq
since the production cross section also decreases.
In fig.~3 we show the cross section for the process
(\ref{eq:pp-h0-gammagamma}) for the case of {\it heavy} squarks.
A characteristic feature of the cross section is that it is small
at moderate values of $m_A$, and then increases steadily with
increasing $m_A$, reaching asymptotically a plateau.
This behaviour is caused by the contribution of the $W$ to the
triangle graph for $h^0\to\gamma\gamma$.
The $h^0WW$ coupling is proportional to $\sin(\beta-\alpha)$,
where
\beq
\cos^2(\beta-\alpha)
=\frac{m_{h^0}^2 (m_Z^2 - m_{h^0}^2)}
      {m_{A^0}^2 (m_{H^0}^2 - m_{h^0}^2)}.
\eeq
For large $m_A$, at fixed $\beta$, all Higgs masses, except $m_{h^0}$,
become large, so that $h^0$ decouples. For large $m_{A}$, we
actually have $\sin(\beta-\alpha)\to1$,
which is why the cross section increases and reaches a plateau
for large $m_{A}$.
\begin{figure}[bht]
\begin{center}
\setlength{\unitlength}{1cm}
\begin{picture}(15.0,5.5)
\put(-0.2,-1)
{\mbox{\epsfysize=5.5cm\epsffile{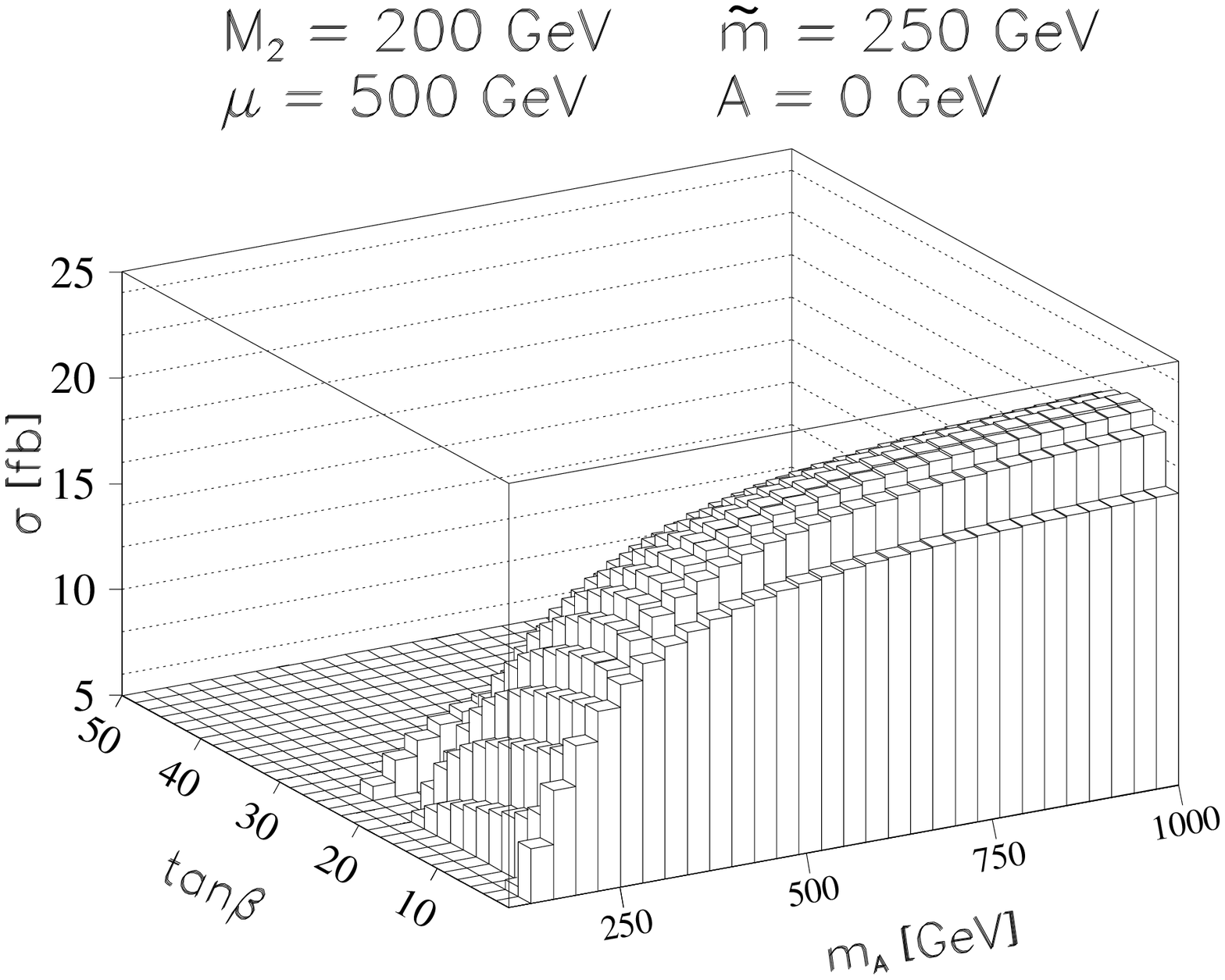}} 
 \mbox{\epsfysize=5.5cm\epsffile{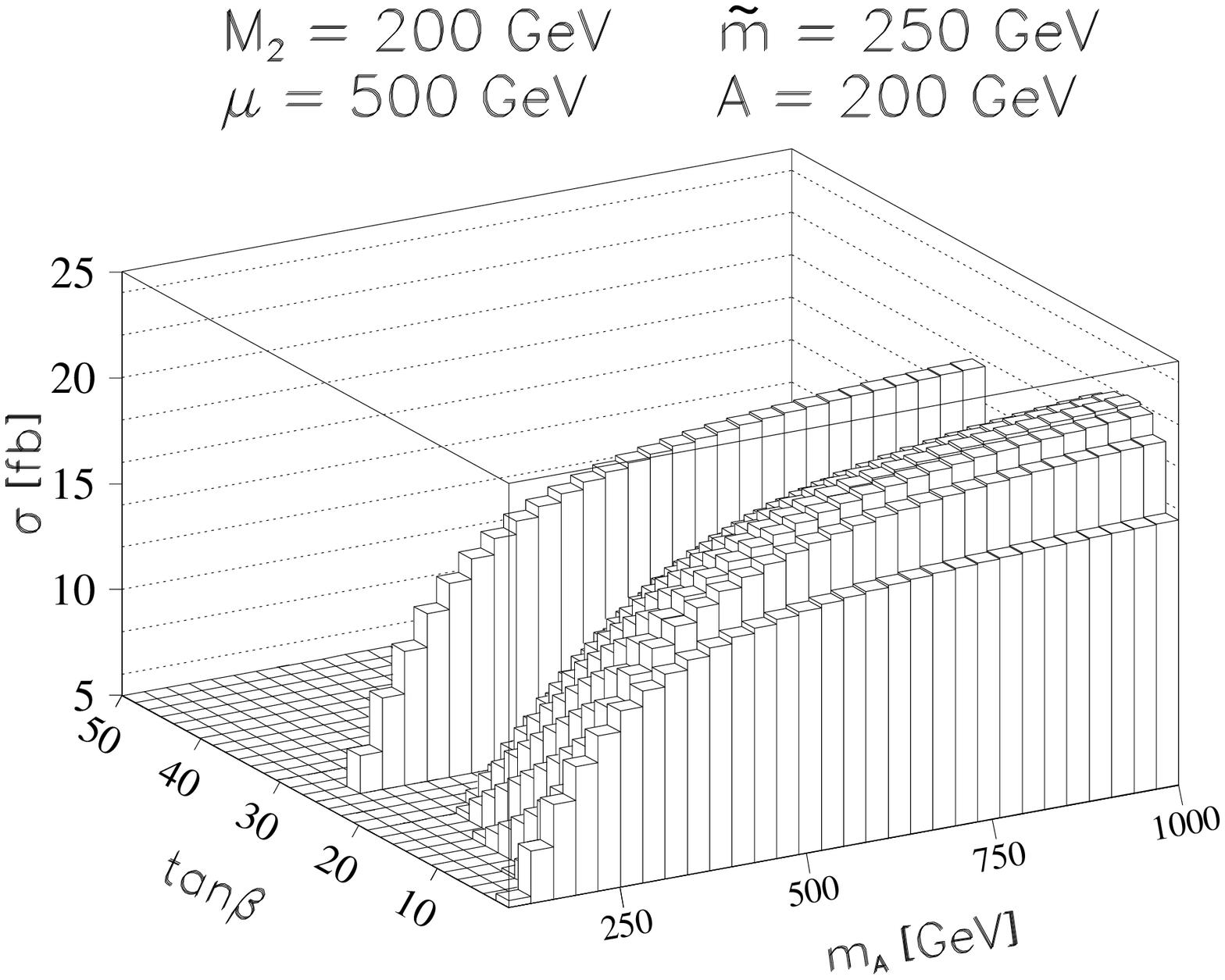}}}
\end{picture}
\begin{capt}
Cross section for $pp\to h^0X\to\gamma\gamma X$ 
as a function of $m_A$ and $\tan\beta$
for $\MNS=200~\GeV$,
$\Ms=250~\GeV$, $\mu=500~\GeV$, and two values of $A$:
{\em left}: $A=0$, {\em right}: $A=200~\GeV$.
\end{capt}
\end{center}
\end{figure}

The situation changes dramatically when we consider lighter squarks.
In fig.~4 we show the corresponding cross section for the case of 
$\Ms=250~\GeV$.
The main difference is that for large values of $\tan\beta$ the
product of the cross section and the two-photon decay rate practically
vanishes.
This is due to the fact that $b$ squarks become light in this limit,
the $h^0$ can decay to $b$ squarks, and the two-photon decay rate 
becomes too small.

For $A=200~\GeV$, there is a band of larger values at 
$\tan\beta\simeq26$. This region is rather ``turbulent"
(for these values of $\Ms$ and $A$):
As mentioned above, the two-photon decay rate vanishes due to
decays to light $b$ squarks,
but at the same time the light Higgs tends to make the cross section big.
The competition between these two effects is responsible for the 
turbulent behaviour seen here.

For some parameters, the latter effect may dominate,
and we get spikes or bands in the
product of the cross section and the two-photon decay rate.
(Since a small Higgs mass is due to large
radiative corrections---which are not accurately known---it is not clear 
how physical these bands or spikes are.)

The $\mu$-dependence of the cross section is for the case of 
$\MNS=200~\GeV$ and $\Ms=500~\GeV$ discussed in refs.~[6,7].
Here, for $\Ms=250~\GeV$, we show in fig.~5 the cross sections
corresponding to those of fig.~4, but for $\mu=-500~\GeV$.
\begin{figure}[thb]
\begin{center}
\setlength{\unitlength}{1cm}
\begin{picture}(15.0,5.5)
\put(-0.2,-1)
{\mbox{\epsfysize=5.5cm\epsffile{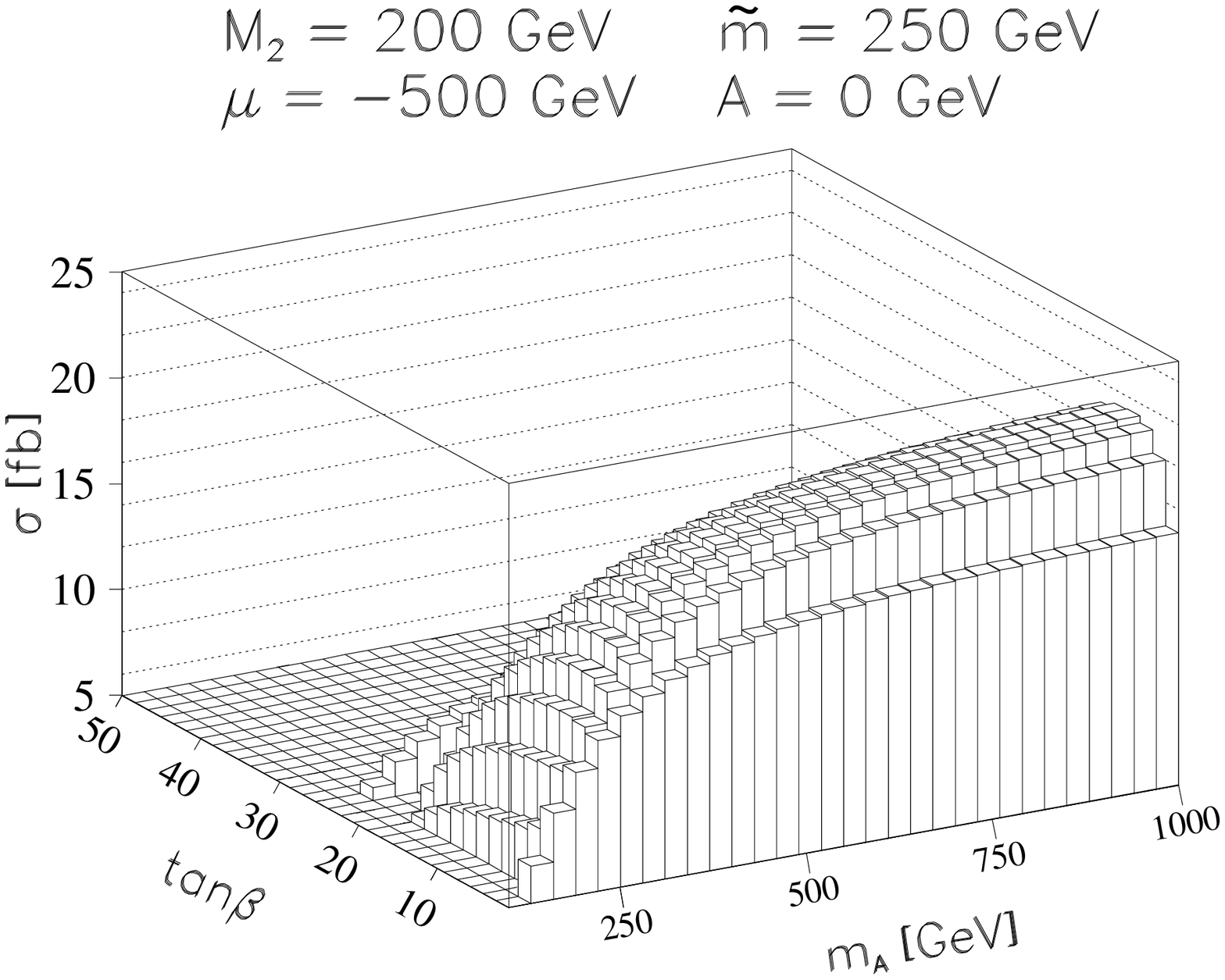}} 
 \mbox{\epsfysize=5.5cm\epsffile{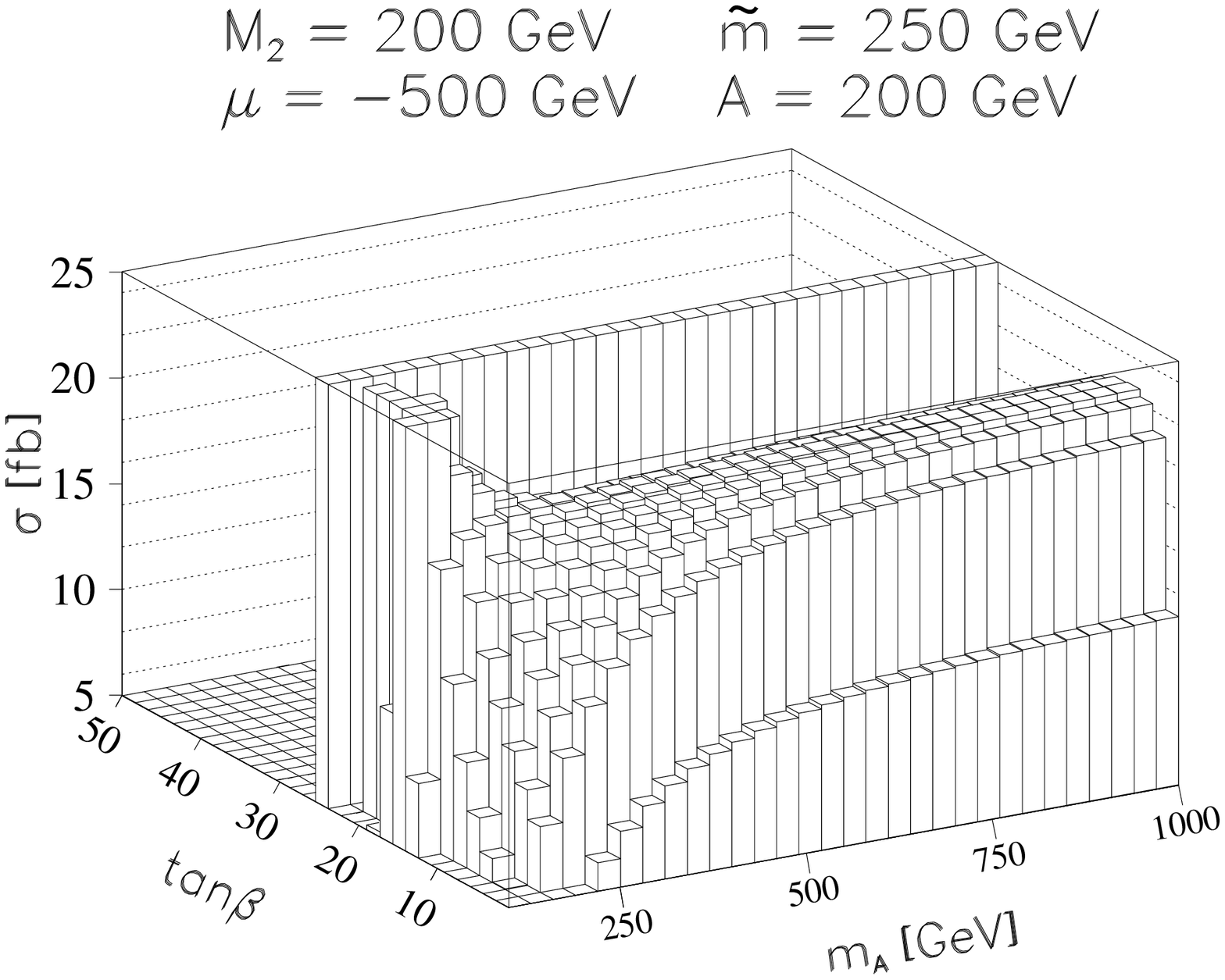}}}
\end{picture}
\begin{capt}
Cross section for $pp\to h^0X\to\gamma\gamma X$ 
as a function of $m_A$ and $\tan\beta$
for $\MNS=200~\GeV$,
$\Ms=250~\GeV$, $\mu=-500~\GeV$, and two values of $A$:
{\em left}: $A=0$, {\em right}: $A=200~\GeV$.
\end{capt}
\end{center}
\end{figure}


\section{Summary and concluding remarks}
\label{sec:conc}
We have discussed the cross section for
the production of the lightest $CP$-even Higgs boson at the LHC, 
in conjunction with its decay to two photons.
Where the parameters lead to a physically acceptable phenomenology,
the cross section multiplied by the two-photon branching ratio
for the lighter $CP$-even Higgs boson is of the order of 15--25~fb.

Similar results have been presented in ref.~[8].
Within the context of a SUGRA GUT model, these authors consider
basically a random sample of parameters compatible with
experimental and theoretical constraints.
The cross sections obtained in ref.~[8] appear to be somewhat
higher than those of ref.~[6].

It should be noted that in regions where the Higgs cross section 
times the two-photon decay rate is small, typically the lightest
$b$ squark is light. 
Thus, as ``compensation", one should be able to observe $b$ squarks.

The recent Fermilab data on large $E_t$ inclusive jet cross 
sections\cite{largeEtjets} suggest that the gluon distribution
functions are larger at high $x$.\cite{cteq4}
We have checked whether these distributions (specifically,
CTEQ4HJ) lead to a higher rate for Higgs production. 
This turns out not to be the case.
The enhancement of the gluon distribution function is in a range 
of $x$ where its magnitude is simply too small, anyway.

These calculations do not take into account QCD corrections.
Such corrections have been evaluated for the quark-loop contribution,
and lead to enhancements of the cross section of about 50\%.\cite{Spira}
For the squark loops QCD corrections have recently been studied
using low-energy theorems.\cite{Dawson} It is concluded that
the additional contributions lead to the same QCD corrections
as for the top and bottom quark loops.

\medskip
It is a pleasure to thank the Organizers of the Minsk Workshop,
in particular Professor L. Tomil'chik,
for creating a very stimulating and pleasant atmosphere during the meeting.
This research has been supported by the Research Council of Norway.
The work of PNP is supported by the Department of Science
and Technology under project No. SP/S2/K-17/94.
                

\end{document}